\documentclass[11pt]{article}
\usepackage{nam,epsfig,graphicx}
\usepackage{subfigure,caption}

\def\Journal#1#2#3#4{{#1} {\bf #2}, #3 (#4)}

\def\JHEP{\em Journal of High Energy Phys.}

\def\NPB{{\em Nucl. Phys.} B}

\def\PRL{\em Phys. Rev. Lett.}
\def\PRD{{\em Phys. Rev.} D}

\def\PPNP{\em Prog. Part. Nucl. Phys.}

\def\ra{\rightarrow}

\def\beq{\begin{equation}}
\def\eeq{\end{equation}}
\def\bea{\begin{eqnarray}}
\def\eea{\end{eqnarray}}
\renewcommand{\d}{\delta}
\renewcommand{\l}{\lambda}

\renewcommand{\b}{\beta}

\newcommand{\m}{\mu}
\newcommand{\E}{{\cal E}}
\renewcommand{\r}{\rho}

\newcommand{\oh}{\frac{1}{2}}

\newcommand{\dg}{\dagger}
\newcommand{\non}{\nonumber}

\newcommand{\pa}{\partial}

\begin{document}
%\vspace*{4cm}
\title{Infrared Divergent Coulomb Self-Energy in
Yang-Mills Theory~\footnote{Talk presented by J.\ Greensite at
{\sl Rencontres du Vietnam V}, Hanoi, Vietnam, August 5-10,
2004.}}

\author{$\underline{\mbox{Jeff Greensite}}{}^\dg$,
{\v S}tefan Olejn\'{\i}k${}^\S$, Daniel Zwanziger${}^\P$}

\address{${}^\dg$Physics and Astronomy Dept., San Francisco State
University, San Francisco, CA 94132 USA \\
${}^\S$Institute of Physics, Slovak Acad. of Sciences,
SK-845 11 Bratislava, Slovakia \\
${}^\P$Physics Dept., New York University, New York, NY 10003 USA}

\maketitle\abstracts{
 It is shown numerically that the Coulomb self-energy
 of an isolated, color non-singlet source diverges in an infinite
 volume.  This is in accord with the Gribov Horizon scenario of
 confinement advocated by Gribov and Zwanziger.  It is also shown
 that this divergence can be attributed to the presence of center
 vortices in thermalized lattice configurations.}

   The energy of an isolated color
non-singlet source, in a confining theory, must be infinite in an infinite volume,
even when the usual ultraviolet divergence is regulated.  We would like to
understand this fact a little better.  Color confinement is usually attributed to center
vortices or abelian monopoles (it is now understood that these objects
are related~\cite{mon-vor}), but there is also another approach to confinement $-$ the
Gribov Horizon scenario, formulated in Coulomb gauge $-$ which has
been advocated by Gribov and Zwanziger.~\cite{GZ}  In this talk I would like
to briefly explain the Gribov Horizon scenario, show numerically that it accounts for the
infrared divergent self-energy of color sources,
and show also that this success is associated with
the presence of center vortices in gauge field configurations.  A more detailed exposition, on
which this talk is based, can be found in a recent article.~\cite{us}

    As a warm-up exercise, let us consider a familiar abelian example:
the electrostatic energy of a static electric point charge, in a cubic volume of
length $L$, with appropriate boundary conditions.  In Coulomb gauge
the Hamiltonian has the form
\bea
       H &=& \oh \int d^3x ~ (E^2 + B^2) ~+~ H_{coul}
~~~,~~~ H_{coul} = \oh \int d^3x d^3y ~ \r(x) K(x,y) \r(y)
\non \\
       K(x,y) &=& \Bigl[M^{-1} (-\nabla^2) M^{-1}\Bigr]_{xy}
~~~~~~~~~~~~~~~~~,~~~ \E = e^2 K(x,x)
\eea
where $\E$ is the Coulomb self-energy of a charge at point $x$.
$M$ is the Faddeev-Popov (FP) operator of the abelian theory.  Let
$A^{\theta}_\m = A_\m - \pa_\m \theta$ be a gauge transformation of
a configuration $A_\m$ fixed to Coulomb gauge.  Then
\beq
      M_{xy} = {\d \over \d \theta(x)} \nabla \cdot A^{\theta}(y)
           = -\nabla^2 \d(x-y)
\eeq
The eigenstates of the FP eigenvalue equation $M \phi^{(n)} =
\lambda_n \phi^{(n)}$ are of course just the plane wave states,
with eigenvalues equal to squared momenta.  In a finite volume
these states are discrete, with a lattice regularization their
number is finite,  and we can express the Green's function
corresponding to $M$ as
\beq
       G_{xy} = \left[ M^{-1} \right]_{xy} = \sum_n {\phi_x^{(n)} \phi_y^{(n)*}
                                                       \over \lambda_n }
\eeq
After some simple manipulations, one finds that
\beq
   {\cal E} = e^2 \Bigl[ M^{-1} (-\nabla^2) M^{-1} \Bigl]_{xx}
     =  {e^2\over L^3} \sum_n  {F(\lambda_n) \over \l_n^2}
\eeq
where $F(\lambda_n) =(\phi^{(n)}| (-\nabla^2)|\phi^{(n)})$, and $L^3$ is
the number of lattice sites in the cubic volume.
Let $\rho(\l)$ denote the normalized density of eigenvalues.
Then at large volumes we can approximate the sum over eigenstates by an integral
\beq
      {\cal E} = e^2 \int d\l ~ { \r(\l) F(\l) \over \l^2}
\label{E}
\eeq
In QED, its easy to show that $\rho(\lambda) = \sqrt{\lambda}/(4 \pi^2)~,~
F(\lambda) = \l$.  The eigenvalues span a finite range, with
$\l_{min} \sim 1/L^2$ and $\l_{max}\sim 1/a^2$, where $L$ is the extension
of the lattice, and $a$ is the lattice spacing.  Putting it all together, we
find that that while $\E$ has an ultraviolet divergence in the continuum
 $a\ra 0$ limit, it is
finite in the infinite volume $L \ra \infty$ limit.  However, finiteness at
infinite volume
clearly depends on the small $\l$ behavior of $\r(\l) F(\l)$.  If instead we had
$\lim_{\l \ra 0} \r(\l) F(\l)/\l > 0$, then the Coulomb energy would be divergent
in the infinite volume limit.

\begin{figure}[htp]
 \centering
    \subfigure[Density $\r(\l)$ of low-lying eigenvalues.]{\label{a1}
         \includegraphics[scale=0.5]{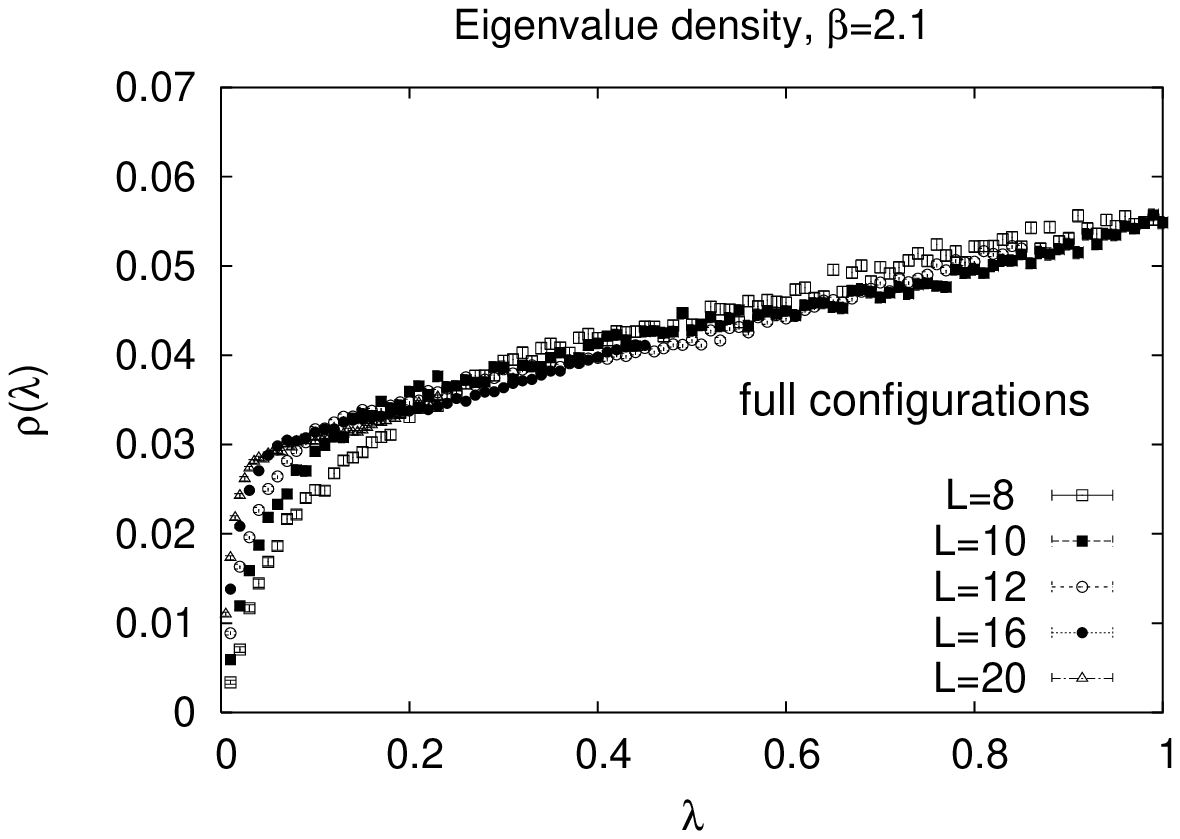}}
          \qquad \qquad
    \subfigure[$F(\l)$ for low-lying eigenstates.]{\label{b1}
         \includegraphics[scale=0.5]{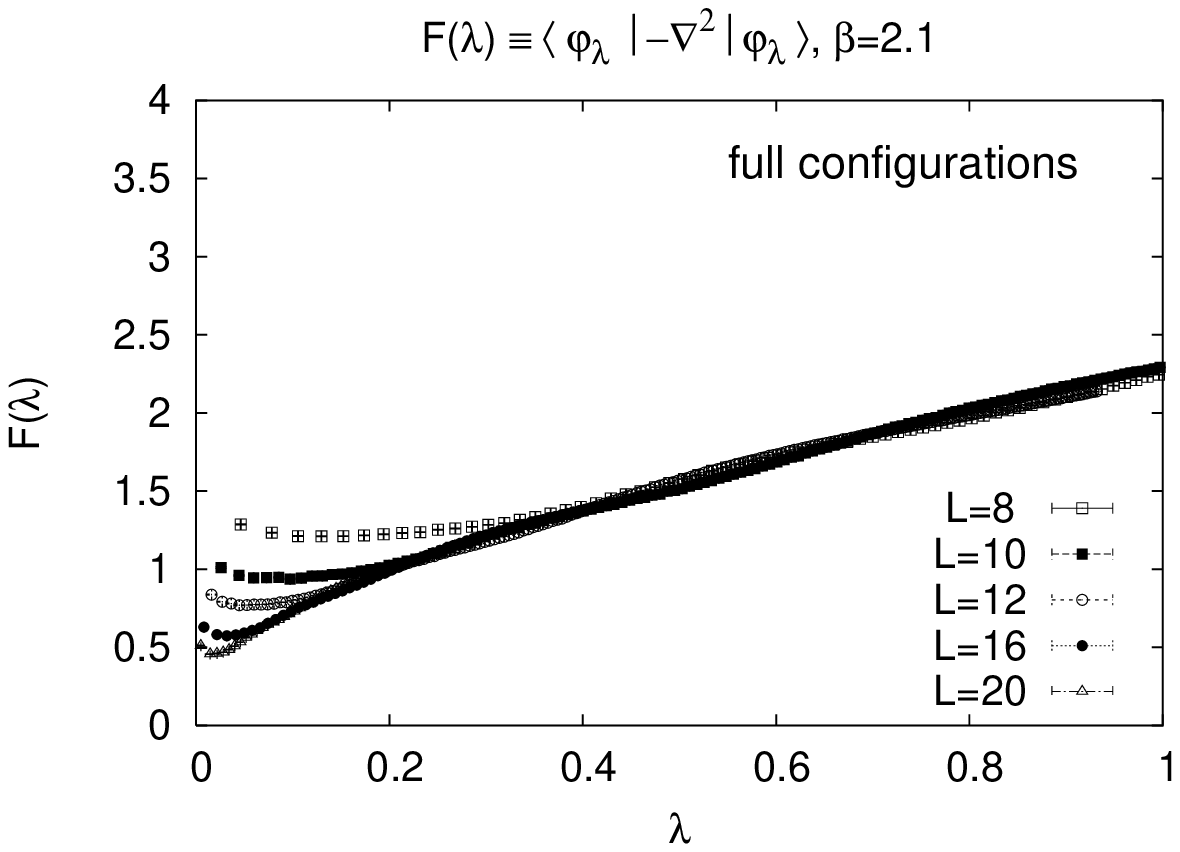}}
  \caption{Data for full, unprojected configurations.}
  \label{fig1}
\end{figure}

In non-abelian theories, there are many gauge copies $-$ Gribov
copies $-$ that satisfy the Coulomb gauge condition.  The Gribov
Region is the space of all Gribov copies with positive FP
eigenvalues.  The boundary of the Gribov region is the
\emph{Gribov Horizon}. Configurations lying on the Gribov Horizon
have at least one FP eigenvalue $\l=0$. Typical Coulomb-gauge
lattice configurations are expected to approach the Gribov horizon,
in the infinite-volume limit, due to entropy considerations.~\cite{Dan}
But what counts for
confinement is the density of eigenvalues $\r(\l)$ near $\l=0$ ,
and the ``smoothness" of these near-zero eigenvalues, as measured
by $F(\l)$.  In non-abelian theories the FP operator is
gauge-field dependent, i.e. $M=-\nabla \cdot D$ where $D_\m$ is the
covariant derivative, and the self-energy of an isolated static
charge in group representation $r$ is proportional to the
quadratic Casimir $C_r$, and to $\E$ in eq.\ (\ref{E}).  So our approach
is to calculate $\r(\l)$ and $F(\l)$ by lattice Monte Carlo
simulations of lattice SU(2) gauge theory, and extrapolate the
results to infinite volume.

   The results are shown in Fig.\ \ref{fig1}, obtained at
$\b=2.1$ for a variety of lattice sizes.  We have applied a scaling
analysis derived from random matrix theory,~\cite{matrix} based on the scaling
of the low-lying eigenvalue distributions with lattice size $L$,
to estimate that in the infinite volume limit we have $\r(\l)\sim
\l^{0.25}, ~ F(\l) \sim \l^{0.4}$ at small $\l$.  Substituting
these power behaviors into eq.\ (\ref{E}), we find that $\E$ has a
divergence in the infrared $L\ra \infty, ~ \l_{min}\ra 0$ limit,
in addition to the usual ultraviolet divergence in the continuum
$a\ra 0$ limit. In other words, the Coulomb self-energy of an
isolated color charge is infrared infinite, by the mechanism
envisaged in the Gribov Horizon scenario.

\begin{figure}[htp]
 \centering
    \subfigure[Density $\r(\l)$ of low-lying eigenvalues.]{\label{a2}
         \includegraphics[scale=0.5]{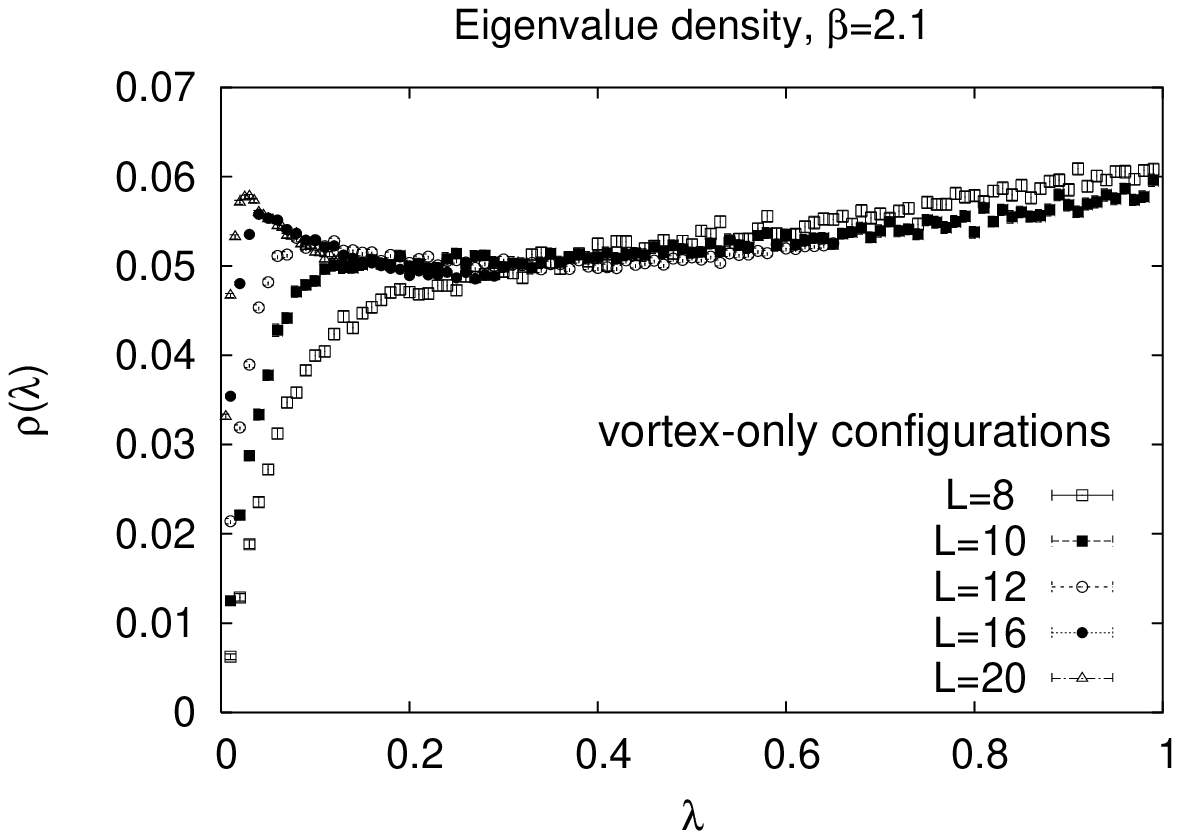}}
          \qquad \qquad
    \subfigure[$F(\l)$ for low-lying eigenstates.]{\label{b2}
         \includegraphics[scale=0.5]{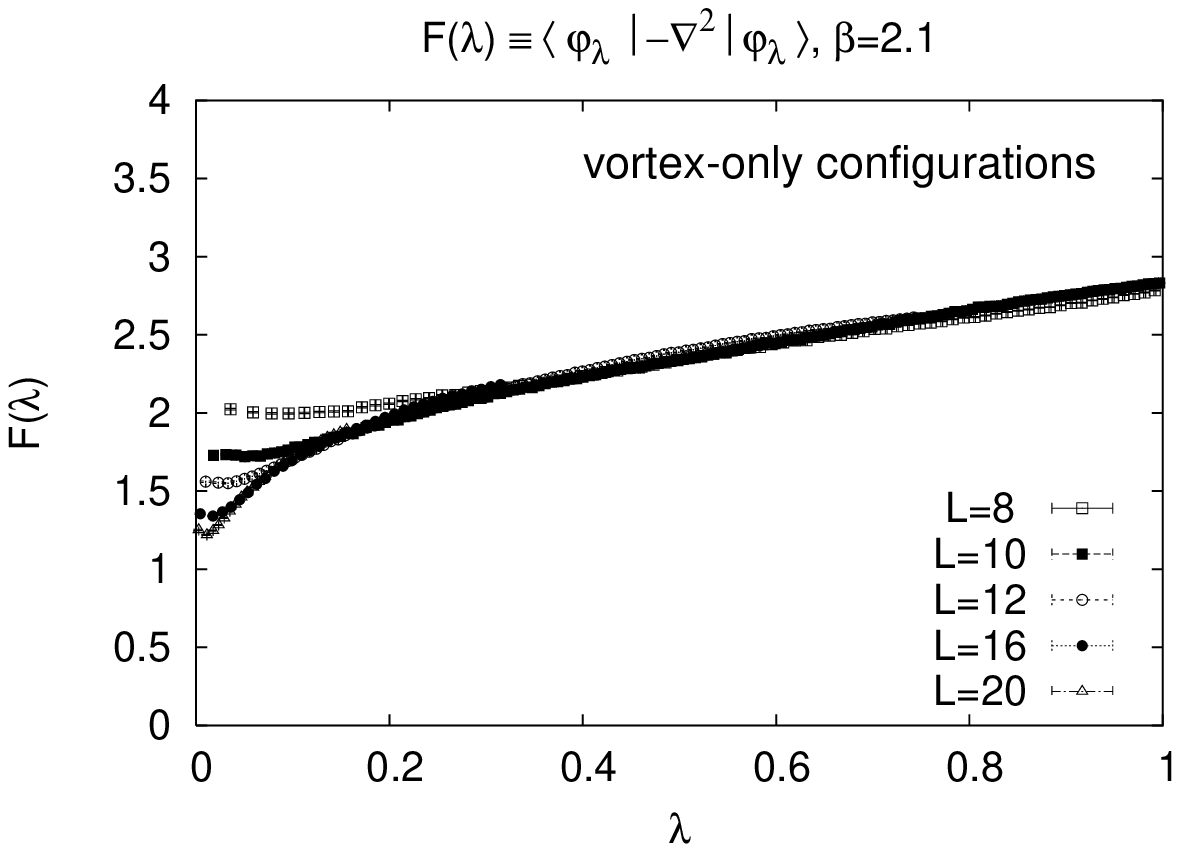}}
  \caption{Data for vortex-only configurations.}
  \label{fig2}
\end{figure}

    We will now display a connection with the center vortex confinement
mechanism.  Recall that center vortices are surfacelike objects in
the (D=4) SU(N) gauge theory vacuum, which can be topologically linked to closed loops.
Creation of a center vortex linked to a Wilson loop multiplies the loop by a center
element of the gauge group.  The center vortex theory of confinement holds
that the area law falloff of Wilson loops is due to vacuum fluctuations in
the number of vortices linking the loop.

    In 1997, methods were devised for locating center vortices in lattice
configurations, and, a little later, for removing them.
This was followed by many investigations of the vortex theory in the lattice community,
with results discussed in two recent reviews.~\cite{Me}  It was found
that (i) center vortices, by themselves, account
for the bulk of asymptotic string tension; (ii) the density of these vortices
scales according to asymptotic freedom; and (iii) removing center vortices
removes the string tension, removes chiral symmetry breaking, and sends
the topological charge to zero.

  We have used the standard technique of maximal center gauge
fixing plus center projection to separate each Monte Carlo
generated configuration into two components: the vortex-only (or ``center-projected")
component, containing only the identified center vortices, and the vortex-removed
component, in which those same vortices have been removed from the original lattice
configuration.  Each component is then transformed to minimal Coulomb gauge.  Our data
for the vortex-only configurations is shown in Fig.\ \ref{fig2}.
This time a finite-volume scaling analysis indicates that
$\rho(\l) \sim \l^{0\pm 0.05},~F(\l) \approx 1$ as $\l\ra 0$,
which again implies (from eq.\ (\ref{E})) an infrared divergence
of the Coulomb energy, resulting from the vortex configurations
alone.

\begin{figure}[htp]
 \centering
    \subfigure[Density $\r(\l)$ of low-lying eigenvalues.]{\label{a3}
         \includegraphics[scale=0.5]{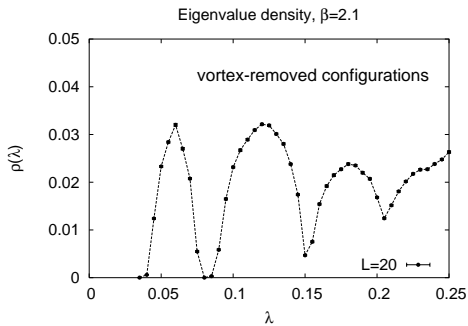}}
         \qquad \qquad
    \subfigure[$F(\l)$ for low-lying eigenstates.]{\label{b3}
         \includegraphics[scale=0.5]{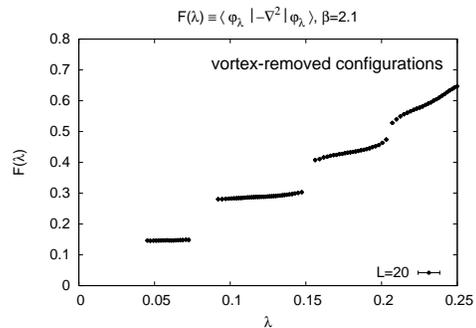}}
  \caption{Data for vortex-removed configurations.}
  \label{fig3}
\end{figure}

    By contrast, when vortices are removed from thermalized
lattice configurations, there is a dramatic change in the
eigenvalue spectrum, illustrated in Fig.\ \ref{fig3} for the
$20^4$ lattice volume.  Inspection of this data reveals that the
number of eigenvalues in each "peak" of $\r(\l)$, and each "band"
of $F(\l)$, matches the degeneracy of eigenvalues of $-\nabla^2$,
the zeroth-order Faddeev-Popov operator, at the given lattice
size. We know that $\r(\l)$ for the $-\nabla^2$ operator at finite
volume is just a
series of $\d$-function peaks.  In the vortex-removed configurations,
these peaks broaden to finite width, but the qualitative features
of $\r(\l), ~ F(\l)$ at zeroth order, i.e.\ the absence of
confinement, remains.

   These numerical results establish that the Coulomb self-energy of
a color non-singlet state is infrared divergent, due to the
enhanced density $\r(\l)$ of Faddeev-Popov eigenvalues near
$\l=0$. This supports the Gribov-Zwanziger picture of confinement.
It also appears that the confining property of the FP eigenvalue
density can be entirely attributed to center vortices, since (i)
enhancement of $\r(\l),~ F(\l)$ is found in the vortex-only content of
lattice gauge configurations; while (ii) the confining properties
of $\r(\l),~ F(\l)$ disappear when vortices are removed from the
lattice.

   We conclude with two further facts about center vortices and
the Gribov horizon, stated  here without proof:  First,
vortex-only configurations have non-trivial Faddeev-Popov zero
modes,~\cite{us1} and therefore lie precisely on the Gribov
horizon, which is a convex manifold in the space of gauge fields,
both in the continuum and on the lattice.  Secondly, vortex-only
configurations are conical singularities on the Gribov
horizon.~\cite{us} It appears that center vortices have a special
geometrical status in Coulomb gauge, although the physical
implications of this fact are not yet fully understood.

\section*{Acknowledgments}
Our research
is supported in part by the U.S. Department of Energy under Grant
No.\ DE-FG03-92ER40711 (J.G.), the Slovak Grant Agency for
Science, Grant No. 2/3106/2003 (\v{S}.O.), and the National Science
Foundation, Grant No. PHY-0099393 (D.Z.).

\end{document}